\documentclass[12pt]{article}
\usepackage[top=25truemm,bottom=27truemm,left=21truemm,right=21truemm]{geometry}

\usepackage{indentfirst}
\usepackage[dvipdfmx]{graphicx}
\usepackage{float}
\usepackage{url}
\usepackage{siunitx}
\usepackage{array}
\usepackage{amsmath}
\usepackage{amssymb}
\usepackage{fancybox}
\usepackage{ascmac}
\usepackage{bm}
\usepackage{mathrsfs}
\usepackage{cancel}
\usepackage{physics}
\usepackage{color}
\usepackage{authblk}
\usepackage{appendix}
\usepackage{comment}
\usepackage{mhchem}
\usepackage{caption,latexsym,amsfonts,mathtools,here}
\usepackage{subcaption}
\usepackage{hyperref}
\usepackage{multirow}
\usepackage[
  backend=biber,
  sorting=none,
  style=numeric-comp,
  doi=false, url=true, eprint=true
]{biblatex}
\numberwithin{equation}{section}

\newcommand{\no}{\nonumber}

\newcommand{\ds}{\displaystyle}

\newcommand{\eqns}{\hspace{-5pt}&=&\hspace{-5pt}}

\newcommand{\simeqs}{\hspace{-5pt}&\simeq&\hspace{-5pt}}
\newcommand{\defs}{\hspace{-5pt}&\equiv&\hspace{-5pt}}
\newcommand{\beqn}{\begin{eqnarray}}
\newcommand{\eeqn}{\end{eqnarray}}

\begin{document}
\begin{titlepage}
\begin{flushright}
{YITP-25-129, KOBE-COSMO-25-16}
\end{flushright}

\vspace{50pt}

\begin{center}

{\large{\textbf{Coherent State Description of Gravitational Waves\\
from Binary Black Holes}}}

\vspace{25pt}

{Sugumi Kanno$^{1,2,3}$, Jiro Soda$^{4}$, and Akira Taniguchi$^1$}
\end{center}

\vspace{20pt}

\shortstack[l]
{\hspace{1.8cm}\it {\small $^1$Department of Physics, Kyushu University, Fukuoka 819-0395, Japan} \\[5pt]
\hspace{1.8cm}\it {\small $^2$Quantum and Spacetime Research Institute, Kyushu University}\\[5pt]
\hspace{1.8cm}\it {\small $^3$Center for Gravitational Physics and Quantum Information,} \\[2pt]
\hspace{1.96cm}\it {\small Yukawa Institute for Theoretical Physics, Kyoto University,} \\[5pt]
\hspace{1.8cm}\it {\small $^4$Department of Physics, Kobe University, Kobe 657-8501, Japan}}
\vspace{50pt}

\begin{abstract}
Quantum mechanics is the fundamental framework of nature, and gravitational waves from binary black holes during the inspiral phase should likewise be analyzed quantum mechanically. It is commonly assumed that their classical description corresponds to a coherent state, so any deviation would signal genuinely quantum nature of gravity. We show that the coherent-state description reproduces classical gravitational waves at leading order, while next-order effects generate squeezed states of gravitons. For GW150914, we estimate the squeezing parameter to be $\sim 10^{-4}$.
We find that gravitational waves from binary black holes are well described by a coherent state.
\end{abstract}
\end{titlepage}
\setcounter{page}{2}
\tableofcontents
\section{Introductioin}

On macroscopic scales, gravity plays a central role. One of the most remarkable predictions of general relativity is the existence of fluctuations in spacetime, known as gravitational waves. In 2015, gravitational waves from a binary black hole merger were directly detected for the first time~\cite{LIGOScientific:2016aoc}.

It is widely accepted that quantum theory provides the fundamental framework governing all physical phenomena at any scale in the universe. Schr{\"o}dinger's cat has become an icon of this view. Upon quantization of gravitational waves, a new particle — the graviton — is predicted. Dyson discussed the detectability of a single graviton and concluded that such detection would be impossible in practice~\cite{Dyson:2013hbl}\footnote{See also Ref.~\cite{Tobar:2023ksi} for a recent proposal of a direct detection method.}. 
However, it has been pointed out that squeezed states may allow for an indirect probe of gravitons~\cite{Parikh:2020nrd,Kanno:2020usf,Parikh:2020kfh,Parikh:2020fhy,Kanno:2021gpt,Ikeda:2025uae} (see also a review \cite{Hsiang:2024qou}).
Thus, it is worthwhile to explore the role of quantum states of gravitons on macroscopic scales\footnote{There are objections to the detection of single gravitons~\cite{Carney:2023nzz,Carney:2024dsj}.}.

From the perspective of the quantum state of gravitons, a key question is how classical gravitational waves should be described within quantum theory. By analogy with quantum optics, it has been implicitly assumed that classical gravitational waves correspond to a coherent state~\cite{Glauber:1963fi}. More recently, this idea has been formalized as the coherent state hypothesis~\cite{Manikandan:2025ykr,Manikandan:2025qgv,Manikandan:2025hlz}. 

In quantum optics, the state of photons is not necessarily a coherent state but can also be a squeezed state, generated through the nonlinear response of a medium. In this sense, photons exhibit distinctly quantum behavior. By analogy, one may expect similar phenomena for gravitons. In particular, squeezing of graviton states can occur in the presence of strong gravitational fields. A well-known example is the squeezed state of primordial gravitational waves generated during inflation~\cite{Grishchuk:1989ss,Grishchuk:1990bj,Albrecht:1992kf}. Another is the squeezed state associated with Hawking radiation from black holes~\cite{Hawking:1975vcx}. Strong gravitational fields are also present in binary systems that emit gravitational waves. Curiously, however, gravitational waves from binary black holes have scarcely been discussed in the context of quantum theory. This omission is often attributed to the macroscopic nature of the system, where quantum effects are usually assumed to be negligible. While this hypothesis is natural, its applicability to such systems where squeezed states may arise due to strong gravity warrants explicit assessment.

For progress in unifying quantum theory and gravity, it is crucial to reveal the non-classical aspects of gravity or, ultimately, to detect gravitons. Recently, the role of nonlinear effects in gravitational waves has been investigated within the framework of quantum theory~\cite{Manikandan:2025dea,Guerreiro:2025sge,Guerreiro:2025mcu}. In addition, squeezed graviton states arising from superradiant axions have been discussed~\cite{Dorlis:2025zzz,Dorlis:2025amf}. 
It should be emphasized, however, that all gravitational waves detected so far originate exclusively from binary black holes. Hence, it is important to investigate the quantum nature of gravitational waves emitted by such systems.
To this end, we first formulate a coherent state description of gravitational waves during the inspiral phase of a binary black hole system. 
We then derive a formula for the squeezing parameters. Furthermore, we estimate the
squeezing parameter for the event GW150914 and discuss the prospects for detecting the quantum nature of gravitational waves, namely gravitons.

\section{Basic framework}
\label{sec:basic}

In the case of photons, a classical current generates a coherent state of photons
as 
\begin{eqnarray}
    \exp[-i \int dt d^3 x  {\bf j}(x^i, t) \cdot {\bf \hat{A}}(x^i,t) ] \ ,
\end{eqnarray}
where $ {\bf j}(x^i , t)$ is a classical current and ${\bf \hat{A}}$ is the operator evaluated along the trajectory. Since the interaction is linear, only a coherent state is produced. 

By contrast, in the case of gravity, a classical source induces the coupling
\begin{eqnarray}
    \exp[-i \int dt d^3 x  \left\{ T_{ij}(x^i, t) \hat{h}_{ij}(x^i,t)
    + \Lambda_{ijkl}(x^i, t) \hat{h}_{ij}(x^i,t) \hat{h}_{kl}(x^i,t)
    + \cdots \right\} ] \ ,
\end{eqnarray}
where $\Lambda_{ijkl}(x^i, t)$ is a classical tensor determined by the interaction between 
matter and gravity. The linear term in $\hat{h}_{ij}$ generates a coherent state, while the quadratic term generates a squeezed state. Our calculation is gauge invariant, since we work in the transverse–traceless gauge and include terms up to second order. In the presence of scalar and vector perturbations, one generally encounters gauge issues due to the mixing between scalar, vector, and tensor modes. However, in our case, we assume a Minkowski background, so no such gauge ambiguity arises. Since the energy flux carried by the created graviton pairs is suppressed by $\sim 1/M_{\rm p}^2$, and the radiation–reaction effect on the binary orbit enters only at order $\left(v/c\right)^5$ according to the quadrupole formula, the backreaction on the background geometry and orbital dynamics can be safely neglected, so that the truncation of the effective gravitational field theory~(EGFT) expansion adopted here is self-consistent.

We consider a binary system with component masses $m_1$ and $m_2$.
The motion of the black holes is described by the geodesics
$\zeta_1 ,\zeta_2$, respectively. For simplicity, 
we assume that the orbital trajectories $x_1^i(t)$ and $x_2^i(t)$ are given, neglecting the back-reaction due to gravitational wave emission. 

The interaction of the corresponding energy-momentum tensor with the graviton field produces a coherent state of gravitons, which describes the classical gravitational waves.
By including higher-order interactions, one can go beyond the coherent-state description. In this section, we derive the interaction Hamiltonian that governs the quantum state of gravitons emitted by the binary black holes.

The total action consists of the Einstein–Hilbert action together with the geodesic actions of the two particles with masses $m_1$ and $m_2$
    \beqn
    \label{eq:action}
	S\eqns S_{\rm EH}+S_{1}+S_{2}=\frac{M_{\rm p}^2}{2}\int d^4x\sqrt{-g}R-m_1\int_{\zeta_1} d\tau-m_2\int_{\zeta_2} d\tau\ ,
	\eeqn
where $M_{\rm p}=1/\sqrt{8\pi G}$ is the Plank mass, $R$ is the Ricci scalar, and $\tau$ is the proper time.

We consider gravitons in the Minkowski space as tensor-mode perturbation of the spatial metric
    \beqn
    \label{eq:metric}
	-d\tau^2 = ds^2=-dt^2 + \left(\delta_{ij}+h_{ij}\right)dx^idx^j\ ,
	\eeqn
where $\delta_{ij}$ is the Kronecker delta and $h_{ij}$ is the metric perturbation, subject to the transverse-traceless conditions $h^i{}_i=h^{i}{}_{j,i}=0$. The indices $(i,j)$ run from 1 to 3, corresponding to  $(x,y,z)$.
 Substituting the metric (\ref{eq:metric}) into the action (\ref{eq:action}) and expanding to second order in $h_{ij}$, we obtain
    \beqn
    \label{eq:free}
	S_{\rm EH}=\frac{M_{\rm p}^2}{8}\int d^4x \left(\dot{h}_{ij}\dot{h}_{ij}-h_{ij,k}h_{ij,k}\right)\ ,
	\eeqn
From this quadratic action, we identify the canonical variable as $\psi_{ij}\equiv h_{ij}M_{\rm p}/2$ with the conjugate momentum given by $\dot{\psi}_{ij}$\,.

We quantize the gravitational waves by imposing the canonical commutation relations,
    \beqn
	\left[\psi_{ij}(t, \bm{x}), \dot{\psi}^{k\ell}(t, \bm{x}')\right] \eqns 
    \frac{i}{2}\left( P_i{}^k P_j{}^\ell
   + P_i{}^\ell P_j{}^k - P_{ij} P^{k\ell}\, \right)
    \delta\left(\bm{x}-\bm{x}'\right)\,,
    \eeqn
where the transverse projection tensor is defined as $P_{ij}=\delta_{ij}-\partial_i \partial_j/\nabla^2$. Thus, the gravitational field $h_{ij}$ can be expanded as follows
 \beqn
    h_{ij}(t,\bm{x})=\frac{2}{M_{\rm p}}\sum_{P=+,\times}\int \frac{d^3\bm{k}}{(2\pi)^{3/2}}\left[\frac{e^{-i\omega_{\bm k} t}}{\sqrt{2\omega_{\bm{k}}}}e^{(P)}_{ij}(\bm{k})a^{(P)}(\bm{k})+\frac{e^{i\omega t}}{\sqrt{2\omega_k}}e^{(P)}_{ij}(-\bm{k})a^{(P)\dag}(-\bm{k})\right]e^{i\bm{k}\cdot\bm{x}}\,,
    \eeqn
where $e^{(P)}_{ij}(\bm{k})\,,(P=+,\times)$ are the polarization tensors, normalized as
$e^{(P)}_{ij}(\bm{k})e^{(Q)}_{ij}(\bm{k})=\delta^{P Q}$.
The anihilation and creation operators satisfy the commutation relation $\left[a^{(P)}(\bm{k}), a^{(Q)\dag}(\bm{k}')\right]=\delta({\bm k}- {\bm k}')\delta^{PQ}$.

The action for the geodesics motion of the particles  $\zeta_N , (N=1,2)$ contains the interaction terms of the form
    \beqn
	S_1+S_2\eqns -\sum_{N=1,2} m_N \int_{\zeta_N} \sqrt{dt^2-\delta_{ij}dx^i dx^j-h_{ij}(t,\bar{\bm{x}}_N (t))dx^i dx^j}\,,\no\\[4pt]
	\eqns -\sum_{N=1,2} m_N\int_{\zeta_N} dt~\frac{1}{\gamma_N}\sqrt{1-\gamma_N^2h_{ij}(t,\bar{\bm{x}}_N (t))v_N^i v_N^j}  \ .
	\eeqn
Here, $\bar{\bm{x}}_N(t)$ denotes the trajectory of the $N$-th particle, $v_N^i=d\bar{x}_N^i/dt$ is its velocity with $v_N^2=v_N^iv_N^i$, and $\gamma_N = 1/\sqrt{1-v_N^2}$ is the Lorentz factor. Therefore,  after performing the Legendre transformation, the interaction Hamiltonian up to second order in $h_{ij}$ takes the form
    \beqn
    \label{eq:intHamiltonian}
	H_{\rm int}(t, \bar{\bm{x}})\eqns \sum_{N=1,2}\left[ 
    \frac{\gamma_N^3 m_N}{2}h_{ij}(t,\bar{\bm{x}}_N(t))v_N^iv_N^j+\frac{3}{8}\gamma_N^5 m_Nh_{ij}(t,\bar{\bm{x}}_N(t))
    \ h_{lm}(t,\bar{\bm{x}}_N(t))\ v_N^iv_N^jv_N^lv_N^m
    \right]\,.\no\\
	\eeqn
We work in the interaction picture, where the time evolution operator $\hat{U}(t,\bar{\bm{x}})$ for the graviton quantum state is governed by the interaction Hamiltonian $\hat{H}_{\rm int}$:
    \beqn
    \label{eq:Uhat}
	\hat{U}(t, \bar{\bm{x}})\eqns \mathcal{T}\left[\exp\left(-i\int^t dt' \hat{H}_{\rm int}(t')\right)\right]
	\eeqn
where $\mathcal{T}$ denotes time ordering.

\section{Coherent state description of gravitational waves}
\label{sec:coherent}

We assume circular orbital motion. Taking the center of mass as the origin, the tragectories can be written as
    \beqn
    \begin{array}{lll}
    x_1= \ds\frac{m_2}{M}a \cos(\Omega t)\ , & y_1 = \ds\frac{m_2}{M}a \sin(\Omega t)\ , & z_1=0\ , \\[10pt]
    x_2= \ds\frac{m_1}{M}a \cos (\Omega t +\pi)\ , & y_2 = \ds\frac{m_1}{M}a \sin(\Omega t +\pi)\ , & z_2=0\ ,
    \end{array}
    \label{trajectories}
    \eeqn
where $M=m_1 + m_2 $ is the total mass and $a$ is the orbital separation. Introducing the reduced mass $\mu=(m_1 m_2) /M$, we obtain the useful relation $m_1 x_1^2 +m_1 y_1^2 + m_2 x_2^2 +m_2 y_2^2 =\mu a^2$. This shows that the reduced mass $\mu$ effectively characterizes the binary system’s quadrupole moment, which plays a central role in the generation of gravitational waves.

The time evolution operator describes the quantum state of gravitons produced by the binary black holes. As discussed in Section~\ref{sec:basic}, retaining only the linear term in $h_{ij}$ in the interaction Hamiltonian yields an operator that generates a coherent state
  \beqn
	\hat{U}(t, \bar{\bm{x}}_N)\eqns \exp\left[-i\frac{2}{M_{\rm p}}\sum_{N=1,2}\frac{\gamma_N^3 m_N}{2}\sum_{P=+,\times}\int^t dt'\int \frac{d^3\bm{k}}{(2\pi)^{3/2}}
  \right. \no\\
    &&\hspace{1.0cm}\left. \times 
    \left(\frac{e^{-i\omega_{\bm k} t'}}{\sqrt{2\omega_{\bm{k}}}}e^{(P)}_{ij}(\bm{k})a(\bm{k})
    +\frac{e^{i\omega_{\bm{k}}t'}}{\sqrt{2\omega_{\bm{k}}}}e^{(P)}_{ij}(-\bm{k})a^\dag(-\bm{k})\right)e^{i\bm{k}\cdot\bar{\bm{x}}_N} v_N^i v_N^j\right]\ .
    \eeqn
Comparing the above expression with the definition of the displacement operator
	\beqn
	\hat{D}(\alpha)=\prod_{P}\exp\left[\int d^3{\bm k} \left(\alpha^{(P)}({\bm k})a^{(P)\dag}({\bm k})-\alpha^{(P)*}({\bm k})a^{(P)}({\bm k})\right)\right]\ ,
	\eeqn    
we identify the coherent state parameter as
    \beqn
    \label{eq:coherentprame}
	\alpha^{(P)}(\bm{k})\eqns -\frac{i}{(2\pi)^{3/2}} \sum_{N=1,2}\int^t dt' \frac{\gamma_N^3 m_N}{M_{\rm p}}\frac{e^{i\omega_{\bm k} t'}}{\sqrt{2\omega_{\bm{k}}}}e^{(P)}_{ij}(\bm{k})v_N^i v_N^j e^{-i\bm{k}\cdot\bar{\bm{x}}_N}\ .
	\eeqn
This parameter encodes the classical orbital dynamics of the binary system into the quantum coherent state of gravitons, thereby providing the bridge between the classical gravitational wave signal and its quantum description.
 
The coherent state $\ket{\alpha}$ is obtained  
by acting with the displacement operator on the vacuum
$\ket{0}$ as $\ket{\alpha} = \hat{D}(\alpha)\ket{0}$.
By definition, it is an eigenstate of the annihilation operator, $a^{(P)}({\bm k})\ket{\alpha}=\alpha^{(P)}({\bm k})\ket{\alpha}$. The expectation value of the metric operator in the coherent state is
    \beqn
    \bra{\alpha} h_{ij}(t,\bm{x})\ket{\alpha}
    =\frac{2}{M_{\rm p}}\sum_{P=+,\times}\int \frac{d^3\bm{k}}{(2\pi)^{3/2}}\frac{e^{(P)}_{ij}(\bm{k})}{\sqrt{2\omega_{\bm{k}}}}
    \left[\alpha^{(P)}(\bm{k})e^{i\bm{k}\cdot\bm{x}-i\omega_{\bm k} t} 
    + \alpha^{(P)*}(\bm{k})e^{-i\bm{k}\cdot\bm{x} +i\omega_k t}\right] \ .
    \eeqn
Choosing the $z$-axis in ${\bm k}$ space to align with the position vector ${\bm x}$, we parametrize the wave vector as $\bm{k}=k(\sin\theta\cos\varphi, \sin\theta\sin\varphi, \cos\theta)$ and substituting the coherent parameters into the above expression, for $r\Omega \gg 1$, we obtain 
\beqn
    \bra{\alpha}h_{xx} \ket{\alpha}
    \eqns \frac{2G \mu(a\Omega)^2}{r}
    \left[ \cos (2\Omega (t+r))  - \cos  (2\Omega (t-r)) \right]
    \ .
\eeqn
where $r=|\bm{x}|$.
Interestingly, this expression contains not only the expected outgoing waves but also ingoing waves. In addition, when compared with the standard quadrupole formula, the amplitude differs by a factor of two. If we add the contributions from both the outgoing and ingoing waves, the correct amplitude is recovered.
It is therefore important to clarify the origin of this apparent discrepancy.

The presence of the ingoing component arises because the coherent-state expectation value incorporates 
all of the directions of wavenumber vectors of the field, corresponding to advanced and retarded solutions of the wave equation. In the standard classical treatment, one imposes retarded boundary conditions to eliminate the ingoing contribution. The discrepancy in amplitude therefore reflects the fact that the coherent-state construction, taken at face value, does not yet enforce the choice of purely retarded (outgoing) solutions.

Now, we can estimate the amplitude of gravitational waves at a distance $r$ from the binary system. As a reference, let us use the parameters of  GW150914~\cite{LIGOScientific:2016aoc}. The component Black hole masses are $36~M_{\odot}$ and $29~M_{\odot}$, giving the reduced mass of $\mu\sim 16~M_{\odot}$. The luminosity distance to the source is $410~{\rm Mpc}$. We consider the orbit of the black holes at the innermost stable circular orbit (ISCO), where the orbital velocity is $a\Omega = 1/\sqrt{6}\sim 0.41$. In this case, the Lorentz factors are $\gamma_1=1.02$ and $\gamma_2=1.03$. Substituting these values, we obtain
    \beqn
   \bra{\alpha} h_{ij}(t,\bm{x})\ket{\alpha}
   \simeq   10^{-21}\left(\frac{\mu}{16~M_{\odot}}\right)\left(\frac{a\Omega}{0.41}\right)^2\left(\frac{410~{\rm Mpc}}{r}\right)
    \ .
    \eeqn
This estimate is consistent with the amplitude of the gravitational waves detected in GW150914.
    
\section{Nonclassicality of gravitational waves}

At the next order of the perturbative expansion in Section~\ref{sec:coherent}, the time evolution operator aquires terms that generates a squeezed state. As discussed in Section~\ref{sec:basic}, such squeezing arises from the term quadratic interaction term proportional to $h_{ij}^2$.  The corresponding operator takes the form 
    \beqn
	\hspace{-0.5cm}U(t, \bar{\bm{x}})
    \eqns \exp\left[-i \sum_{N=1,2} \int^t dt' \frac{3\gamma_N^5 m_N}{2M_{\rm p}^2}\sum_{P,Q}\int \frac{d^3{\bm k}}{(2\pi)^{3/2}} \int \frac{d^3{\bm k}'}{(2\pi)^{3/2}}\right.\no\\[6pt]
    &&\hspace{-0.5cm}\left.\times\left(\frac{e^{-i(\omega_k+\omega_{k'}) t'}}{2\sqrt{\omega_k\omega_{k'}}}
    e^{(P)}_{ij}({\bm k})e^{(Q)}_{lm}({\bm k}')
    v_N^i v_N^j v_N^l v_N^m a^{(P)}({\bm k}) a^{(Q)}({\bm k}')e^{i\bm{k}\cdot\bar{\bm{x}}_N}e^{i\bm{k}'\cdot\bar{\bm{x}}_N}+{\rm h.c.}\right)\right]\ .
    \label{eq:sqevo}
	\eeqn
The terms proportional to $aa^\dag$ or $a^\dag a$ correspond only to phase rotations and do not contribute to the squeezing amplitude. We therefore neglect them. 

Comparing this operator with the definition of the squeeze operator,
	\beqn
	\hat{S}(\beta)=\prod_{P}\exp\left[\int d^3{\bm k}\int d^3{\bm k}' \left(\beta^{(P)}_{kk'}a^{(P)\dag}({\bm k})a^{(P)\dag}({\bm k}')-\beta_{kk'}^{(P)*}a^{(P)}({\bm k}) a^{(P)}({\bm k}')\right)\right]\ ,\label{eq:squeezeope}
	\eeqn
we can identify the squeezing parameter as
    \beqn  
	\hspace{-1.0cm}\beta^{(P)}_{kk'}\eqns -\frac{3i}{(2\pi)^3 2M_{\rm p}^2}\sum_{N=1,2} 
    \gamma_N^5 m_N\int^t dt'
    \frac{e^{i(\omega_k+\omega_{k'}) t'}}{2\sqrt{\omega_k\omega_{k'}}}
    e^{(P)}_{ij}({\bm k})e^{(P)}_{lm}({\bm k}')
    v_N^i v_N^j v_N^l v_N^m e^{-i\bm{k}\cdot\bar{\bm{x}}_N}e^{-i\bm{k}'\cdot\bar{\bm{x}}_N}\ .
    \label{eq:beta}
	\eeqn
Here, after azimuthal integration over ${\bm k}$ and ${\bm k}'$, the cross terms between the $+$ and $\times$ polarizations vanish because their angular dependences (proportional to $\cos 2\varphi$ and $\sin 2\varphi$) are orthogonal, so that only the diagonal ($P=Q$) contributions remain.
Constraints on the squeeing parameters have been discussed in the literature~\cite{Hertzberg:2021rbl}, based on LIGO data~\cite{McCuller:2021mbn}. 

We now evaluate the squeezing parameter at a location far from the binary system. The expressions derived above correspond to the values at the source. In our analysis we assume that environment-induced decoherence of the squeezed graviton state between emission and observation is negligible. This assumption can be justified as follows. For a typical LIGO-band frequency $f \sim 100~\mathrm{Hz}$, the gravitational-wave wavelength is $\lambda \sim 10^{8}\,\mathrm{cm}$, so dust grains and other small-scale inhomogeneities are deep in the long-wavelength regime, where their effect on the wave is strongly suppressed by the equivalence principle. In addition, the graviton–matter scattering cross section is of order the Planck area, $L_{\rm Pl}^2 \sim 10^{-65}\,\mathrm{cm}^2$~\cite{Dyson:2013hbl}, which leads to an extremely small interaction probability for gravitons propagating through the interstellar and intergalactic medium. Consequently, the optical depth for graviton scattering is $\tau \ll 1$, and environment-induced decoherence of the squeezed state during propagation from the binary to the detector can safely be neglected. Under this assumption, the value of the squeezing parameter at the detector is identical to that at the source.

By performing the time integration, we see that the squeezing parameter attains its largest value when ${\bm k} \simeq -{\bm k}'$ and $\omega_k = \omega_{k'} \simeq 2\Omega$. The first relation reflects total momentum conservation of the emitted graviton pair, and the second corresponds to the quadrupole frequency of the binary. The squeezing parameter at the location of the binary system can be estimated as
    \beqn
    \label{eq:squzeta}
    \zeta \simeq \frac{4\pi}{3} (2\Omega)^3|\beta|
    \simeq \frac{1}{8\pi M_{\rm p}^2}\mu (a\Omega)^4 f \ .
    \eeqn
In the second expression, we have used the relation $\omega=2\pi f$. As Eq.~(\ref{eq:squzeta}) shows, the factor $1/M_{\rm p}^2$ represents the usual Planck-suppressed gravitational coupling. For stellar-mass binary black holes, however, the reduced mass is macroscopic, $\mu \sim 10\,M_\odot \sim 10^{34}\,\mathrm{g}$, while the Planck mass is $M_{\rm p} \sim 10^{-5}\,\mathrm{g}$, so that $\mu/M_{\rm p} \sim 10^{39}$. This enormous hierarchy illustrates how the macroscopic source can compensate much of the Planck suppression. As in Section~\ref{sec:coherent}, we adopt the parameters of  GW150914~\cite{LIGOScientific:2016aoc}. In this case, we obtain
    \beqn
    \label{eq:sqpara}
    \zeta \simeq 2\times 10^{-4}\left(\frac{\mu}{16~M_\odot}\right)\left(\frac{a\Omega}{0.41}\right)^4\left(\frac{f}{68~{\rm Hz}}\right)\ .
    \eeqn
Thus, the quantum state of gravitational waves from GW150914 is characterized by a squeezing parameter of order $10^{-4}$. Note that the squeezing parameter in Eq.~(\ref{eq:sqpara}) depends only weakly on the black-hole masses. This is because the reduced  mass $\mu$ and the characteristic angular frequency of the emitted gravitational waves $\Omega$ are related (more massive binaries radiate at lower frequencies). While this estimate is based on the ISCO, the squeezing is expected to be even stronger just before the merger. For example, in GW150914 the frequency of the gravitational wave reached up to $150~{\rm Hz}$.
\section{Conclusion}

In this work, we investigated how astrophysical binary black holes generate a quantum state of gravitons. We modeled the binary system during the inspiral phase. as a classical source. As is well known, when a classical source couples linearly to a quantum field, it produces a coherent state.

We confirmed that classical gravitational waves can be described within the coherent state framework. We then extended the analysis to the next-order effects and showed that binary black holes can also generate squeezed states. Furthermore, we estimated the degree of squeezing and found that, for GW150914, the squeezing parameter is of order $10^{-4}$. Note that this value does not depend much on the black hole mass. This result indicates that although the squeezed components can theoretically occur at the second order due to nonlinear effects, a coherent state well approximates the quantum state of gravitational waves from binary black holes.

Since we have obtained a squeezed coherent state, Hanbury Brown-Twiss interferometry may be employed to probe the nonclassicality of the graviton state~\cite{HanburyBrown:1956bqd,Brown:1956zza,Kanno:2018cuk}
with new technology~\cite{Vermeulen:2024vgl,Patra:2024eke}.
It would also be interesting to extend our analysis to other sources of gravitational waves. In this paper, we have demonstrated how a quantum state is generated by  binary black holes under the assumption of the standard vacuum. However, if primordial gravitational waves generated during inflation are taken into account, an additional enhancement of squeezing is expected. Thus, by observing the squeezing of gravitational waves from binary black holes, one may obtain indirect information about the early universe. For details, see the reference~\cite{Kanno:2025fpz}.

\section*{Acknowledgments}
S.\ K. was supported by the Japan Society for the Promotion of Science (JSPS) KAKENHI Grant Numbers JP22H01220, 24K21548 and MEXT KAKENHI Grant-in-Aid for Transformative Research Areas A “Extreme Universe” No. 24H00967.
J.\ S. was in part supported by JSPS KAKENHI Grant Numbers JP23K22491, JP24K21548, JP25H02186. A.\ T. was supported by JSPS KAKENHI Grant Number JP25KJ1912.
\appendix
\section{Derivation of the coherent parameter}
The coherent parameter is expressed as
    \beqn
	\alpha^{(P)}(\bm{k})\eqns -\frac{i}{(2\pi)^{3/2}} \sum_{N=1,2}\int^t dt' \frac{\gamma_N^3 m_N}{M_{\rm p}}\frac{e^{i\omega_{\bm k} t'}}{\sqrt{2\omega_{\bm{k}}}}e^{(P)}_{ij}(\bm{k})v_N^i v_N^j e^{-i\bm{k}\cdot\bar{\bm{x}}_N}\ .
	\eeqn
Choosing the $z$-axis in ${\bm k}$ space to align with the position vector ${\bm x}$, we parametrize the wave vector as $\bm{k}=k(\sin\theta\cos\varphi, \sin\theta\sin\varphi, \cos\theta)$. The coherent state parameter is specifically written as
    \beqn
    \alpha^{(+)}(\bm{k})
    \eqns \frac{i}{(2\pi)^{3/2}} \frac{\mu (a\Omega)^2}{\sqrt{2} M_{\rm p}}\int^t dt' \frac{e^{i\omega_{\bm k} t'}}{\sqrt{2\omega_{\bm{k}}}}\left(\frac{\sin^2\theta}{2}+\frac{1+\cos^2\theta}{2}\cos(2\Omega t'- 2\varphi)\right)\no\\[6pt]
    &&\hspace{3.5cm} \times\left[\gamma^3_1 \left(\frac{m_2}{M}\right)e^{-i(k_xx_1+k_yy_1)}+\gamma^3_2 \left(\frac{m_1}{M}\right)e^{-i(k_xx_2+k_yy_2)}\right]\ ,\label{eq:parameterplus}\\[10pt]
    \alpha^{(\times)}(\bm{k})
    \eqns \frac{i}{(2\pi)^{3/2}} \frac{\mu (a\Omega)^2}{\sqrt{2}M_{\rm p}}\int^t dt' \frac{e^{i\omega_{\bm k} t'}}{\sqrt{2\omega_{\bm{k}}}}\cos\theta\sin(2\Omega t' - 2\varphi)\no\\[6pt]
    &&\hspace{3.5cm} \times \left[\gamma^3_1 \left(\frac{m_2}{M}\right)e^{-i(k_xx_1+k_yy_1)}+\gamma^3_2 \left(\frac{m_1}{M}\right)e^{-i(k_xx_2+k_yy_2)}\right]\ ,\label{eq:parametercross}
    \eeqn
where 
    \beqn
    k_xx_1+k_yy_1\eqns \frac{m_2}{M}ak(\sin\theta\cos\varphi \cos(\Omega t')+\sin\theta\sin\varphi \sin(\Omega t'))\ ,\\[6pt]
    k_xx_2+k_yy_2\eqns -\frac{m_1}{M}ak(\sin\theta\cos\varphi \cos(\Omega t')+\sin\theta\sin\varphi \sin(\Omega t'))\ .
    \eeqn
We also used the explicit form of polarization tensors 
 \begin{eqnarray}
    e^{(+)}_{ij}\eqns \frac{1}{\sqrt{2}}
    \begin{pmatrix}
    \cos^2\theta \cos^2\varphi-\sin^2\varphi & (1+\cos^2\theta)\sin\varphi\cos\varphi & -\ds\frac{1}{2}\sin 2\theta \cos\varphi \\[12pt]
    (1+\cos^2\theta)\sin\varphi\cos\varphi & \cos^2\theta \sin^2\varphi-\cos^2\varphi & -\ds\frac{1}{2}\sin 2\theta \sin\varphi \\[12pt]
    -\ds\frac{1}{2}\sin 2\theta \cos\varphi & -\ds\frac{1}{2}\sin 2\theta \sin\varphi & \sin^2\theta
    \label{eij-plus}
    \end{pmatrix},\\[12pt] 
    e^{(\times)}_{ij}\eqns \frac{1}{\sqrt{2}}
    \begin{pmatrix}
    -\cos\theta \sin 2\varphi & \cos\theta \cos 2\varphi & \sin\theta \sin\varphi \\[12pt]
    \cos\theta \cos 2\varphi & \cos\theta \sin 2\varphi & -\sin\theta \cos\varphi \\[12pt]
    \sin\theta \sin\varphi & -\sin\theta \cos\varphi & 0
    \label{eij-cross}
    \end{pmatrix}
    \ .
    \end{eqnarray}
The expectation value of the metric operator in the coherent state is
    \beqn
    \bra{\alpha} h_{ij}(t,\bm{x})\ket{\alpha}
    =\frac{2}{M_{\rm p}}\sum_{P=+,\times}\int \frac{d^3\bm{k}}{(2\pi)^{3/2}}\frac{e^{(P)}_{ij}(\bm{k})}{\sqrt{2\omega_{\bm{k}}}}
    \left[\alpha^{(P)}(\bm{k})e^{i\bm{k}\cdot\bm{x}-i\omega_{\bm k} t} 
    + \alpha^{(P)*}(\bm{k})e^{-i\bm{k}\cdot\bm{x} +i\omega_k t}\right] \ .
    \eeqn
As an example, we show the calculation of the $h_{xx}$ component.
    \beqn
    \label{eq:ahxxa}
    \bra{\alpha}h_{xx}\ket{\alpha}= \frac{\mu(a\Omega)^2}{\sqrt{2}\pi M_{\rm p}^2}\frac{1}{r} \left[F^{(+)}(t, r)+F^{(\times)}(t, r)\right]\ ,
    \eeqn
where we defined $F^{(+)}(t, r)$ and $F^{(\times)}(t, r)$ as follows.
    \beqn
    F^{(+)}(t, r)\defs \int_0^\infty r dk ~\frac{1}{2}\int_0^\pi \sin\theta d\theta ~\frac{1}{2\pi}\int_0^{2\pi} d\varphi~\frac{1}{2\pi}\int_{-\infty}^t k dt'\no\\[6pt]
    &&\times \left\{i\left[\gamma^3_1 \left(\frac{m_2}{M}\right) e^{-i(k_xx_1+k_yy_1)}+\gamma^3_2 \left(\frac{m_1}{M}\right) e^{-i(k_xx_2+k_yy_2)}\right]e^{ikr\cos\theta-i\omega_{k} t} e^{i\omega_{k} t'}\right.\no\\[6pt]
    &&\hspace{24pt}\left. \times e_{xx}^{(+)}\left(\frac{\sin^2\theta}{2}+\frac{1+\cos^2\theta}{2}\cos(2\Omega t'-2\varphi)\right)+{\rm c.c.}\right\}\, ,\\[6pt]
    F^{(\times)}(t, r)\defs\int_0^\infty r dk ~\frac{1}{2}\int_0^\pi \sin\theta d\theta ~\frac{1}{2\pi}\int_0^{2\pi} d\varphi~\frac{1}{2\pi}\int_{-\infty}^t k dt'\no\\[6pt]
    &&\times \left\{i\left[\gamma^3_1 \left(\frac{m_2}{M}\right) e^{-i(k_xx_1+k_yy_1)}+\gamma^3_2 \left(\frac{m_1}{M}\right) e^{-i(k_xx_2+k_yy_2)}\right]e^{ikr\cos\theta-i\omega_{k} t} e^{i\omega_{k} t'}\right.\no\\[6pt]
    &&\hspace{24pt} \left. \times e_{xx}^{(\times)}\cos\theta\sin\left(2\Omega t'-2 \varphi\right)+{\rm c.c.}\right\}\ ,
    \eeqn
where $r=|\bm{x}|$ and the integrals can be evaluated approximately. Next we assume that a binary black hole system with two masses of approximately equal, and set
    \beqn
    \gamma^3_1 \left(\frac{m_2}{M}\right) e^{-i(k_xx_1+k_yy_1)}+\gamma^3_2 \left(\frac{m_1}{M}\right) e^{-i(k_xx_2+k_yy_2)}\simeq 1\ .
    \eeqn
Under this assumption, we obtain
    \beqn
    &&\hspace{-12pt}F^{(+)}(t, r)\no\\[6pt]
    \eqns \int_0^\infty r dk ~\frac{1}{2}\int_0^\pi \sin\theta d\theta ~\frac{1}{2\pi}\int_0^{2\pi} d\varphi~\frac{1}{2\pi}\int_{-\infty}^t k dt'\no\\[6pt]
    &&\left\{\frac{i}{2}e^{ikr\cos\theta-i\omega_{k}t}\left[e_{xx}^{(+)}\sin^2\theta e^{i\omega_{k} t'}+e_{xx}^{(+)}\frac{1+\cos^2\theta}{2}e^{i\omega_k t'}\left(e^{i2\Omega t'}e^{-2i\varphi}+e^{-i2\Omega t'}e^{2i\varphi}\right)\right]+{\rm c.c.}\right\}\, .\no\\
\eeqn
When $\Omega>0$, the second term oscillates more rapidly than the third term, so the second term can be neglected. When $\Omega<0$, the situation is reversed, and the third term can be neglected. Now we take $\Omega>0$ and performe the integration, we have
    \begin{eqnarray}
    &&\hspace{-12pt}F^{(+)}(t,r)\no\\[6pt]
    \eqns r\int_0^\infty dk~\frac{1}{2}\int_0^\pi \sin\theta d\theta ~\frac{1}{2\pi}\int_0^{2\pi} d\varphi\no\\[6pt]
    && k \left\{\frac{i}{2}e^{ikr\cos\theta-i\omega_{k}t}\left[e_{xx}^{(+)}\sin^2\theta\delta(k-0)+e_{xx}^{(+)}\frac{1+\cos^2\theta}{2}\delta(k-2\Omega)e^{2i\varphi}\right]+{\rm c.c.}\right\}\no\\[6pt]
    \eqns \frac{r~2\Omega}{8\pi}\int_0^\pi \sin\theta d\theta \int_0^{2\pi} d\varphi \left\{ie^{i2\Omega(r\cos\theta-t)}\frac{1}{\sqrt{2}}\left(\cos^2\theta \cos^2\varphi-\sin^2\varphi\right)\frac{1+\cos^2\theta}{2}e^{2i\varphi}+{\rm c.c.}\right\}\no\\[6pt]
    \eqns \frac{\sqrt{2}}{64(r\Omega)^4}\left\{i e^{-i2\Omega t}\left[2r\Omega\left(-3+4(r\Omega)^2\right)\cos(2r\Omega)+\left(3-8(r\Omega)^2+8(r\Omega)^4\sin(2r\Omega)\right) \right]+{\rm c.c.}\right\}\ .\no\\
    \end{eqnarray}
Since the distance from a binary black holes $r$ is enough larger than the Schwarzschild radius, we can take $r\Omega \gg 1$, and then obtain
    \beqn
    \label{eq:Fplus}
    F^{(+)}(t,r) \simeqs \frac{\sqrt{2}}{64(r\Omega)^4}\left\{i e^{i2\Omega t}\left[8(r\Omega)^4\sin(2r\Omega)\right]+{\rm c.c.}\right\}\no\\[6pt]
    \eqns \frac{1}{4\sqrt{2}}\left[\cos(2\Omega(t+r))-\cos(2\Omega(t-r))\right]\ .
    \eeqn
Similarly, we can calculate $F^{(\times)}(t,r)$ as well.
    \beqn
    \label{eq:Fcross}
    F^{(\times)}(t,r) \eqns \frac{1}{4\sqrt{2}}\left[\cos(2\Omega(t+r))-\cos(2\Omega(t-r))\right]\ .
    \eeqn
Thus, substituting Eq.~(\ref{eq:Fplus}) and (\ref{eq:Fcross}) into Eq.~(\ref{eq:ahxxa}), we obtain
    \beqn
    \bra{\alpha}h_{xx}\ket{\alpha}\eqns \frac{\mu(a\Omega)^2}{4\pi M_{\rm p}^2}\frac{1}{r} \left[\cos(2\Omega(t+r))-\cos(2\Omega(t-r))\right]\no\\[6pt]
    \eqns \frac{2G\mu(a\Omega)^2}{r} \left[\cos(2\Omega(t+r))-\cos(2\Omega(t-r))\right]\, .
    \eeqn
\section{Derivation of the squeezing parameter}
The squeezing parameter is expressed as
    \beqn  
	\hspace{-1.0cm}\beta^{(PQ)}_{kk'}\eqns -\frac{3i}{(2\pi)^3 2M_{\rm p}^2}\sum_{N=1,2} 
    \gamma_N^5 m_N\int^t dt'
    \frac{e^{i(\omega_k+\omega_{k'}) t'}}{2\sqrt{\omega_k\omega_{k'}}}
    e^{(P)}_{ij}({\bm k})e^{(Q)}_{lm}({\bm k}')
    v_N^i v_N^j v_N^l v_N^m e^{-i\bm{k}\cdot\bar{\bm{x}}_N}e^{-i\bm{k}'\cdot\bar{\bm{x}}_N}\ .
    \label{eq:beta}
	\eeqn
Since Eq.~(\ref{eq:beta}) contains the phase factor $\exp\left\{-i\left[(\bm{k}+\bm{k}')\cdot \bar{\bm{x}}_N-(\omega_k+\omega_{k'})t'\right]\right\}$, by performing the integration with respect to time, we obtain relations $\bm{k}+\bm{k}'=0$ and $\omega_k=k=2\Omega$. Physically, the first condition reflects conservation of the total momentum of the emitted pair due to translational symmetry about the binary’s center of mass, and the second frequency selection is the same as that given by the classical quadrupole formula. After performing azimuthal integrating of $\bm{k}, -\bm{k}$, namely $(\theta',\varphi')=(\pi-\theta, \varphi+2\pi)$, the cross terms involving $+$ and $\times$ modes vanish.

In Eq.~(B.1), the kernel involves the tensor contraction
\[
e^{(P)}_{ij}({\bm k})\,e^{(Q)}_{lm}({\bm k}')\,v_N^i v_N^j v_N^l v_N^m \, .
\]
For a given wavevector ${\bm k}$, the projection of the polarization tensors on the orbital velocity can be written schematically as
\[
e^{(+)}_{ij}({\bm k}) v_N^i v_N^j \propto F_+(\theta)\cos\bigl(2\Omega t' - 2\varphi\bigr), \qquad
e^{(\times)}_{ij}({\bm k}) v_N^i v_N^j \propto F_\times(\theta)\sin\bigl(2\Omega t' - 2\varphi\bigr),
\]
with some functions $F_+(\theta)$ and $F_\times(\theta)$ of the polar angle $\theta$. Thus the cross term with $P \neq Q$ contains a product of the form
\[
\cos\bigl(2\Omega t' - 2\varphi\bigr)\,\sin\bigl(2\Omega t' - 2\varphi\bigr),
\]
whose integral over the azimuthal angle $\varphi$ vanishes. More explicitly, we find
\begin{eqnarray}
&&\hspace{-12pt}\int_0^\pi d\theta \sin \theta \int_0^{2\pi} d\varphi\;
e^{(+)}_{ij}({\bm k})e^{(\times)}_{lm}(-{\bm k})
v_1^i v_1^j v_1^l v_1^m\,
e^{-i{\bm k}\cdot\bar{\bm x}_N}e^{-i(-{\bm k})\cdot\bar{\bm x}_N} \nonumber \\[6pt]
&=& \frac{1}{2}\left(\frac{m_2}{M}a\Omega\right)^4
\int_0^\pi d\theta \sin \theta \int_0^{2\pi} d\varphi \nonumber \\[6pt]
&&\times \left[\frac{\sin^2\theta}{2}
+ \frac{1+\cos^2\theta}{2}\cos(2\Omega t' - 2\varphi)\right]
\cos(\pi-\theta)\,
\sin\!\left(2\Omega t' - 2(\varphi+2\pi)\right) \nonumber \\[6pt]
&=& 0 \, .
\end{eqnarray}
This shows that, after integration over the azimuthal angles of ${\bm k}$ and ${\bm k}'$, the cross terms with $P \neq Q$ vanish, and only the diagonal contributions with $P = Q$ survive. By contrast, in the circular (helicity) basis defined by
$e_{ij}^{\mathrm{R,L}} = (e_{ij}^{(+)} \pm i\,e_{ij}^{(\times)})/\sqrt{2}$, the kernel generally acquires off–diagonal components because the diagonal components are not equal for our binary configuration. Thus, right- and left-handed polarizations are correlated in this basis. 
\\

Therefore, by using Eqs.~(\ref{eij-plus}) and (\ref{eij-cross}), the squeezing parameter can be written explicitly as
    \beqn
    \beta^{(+)}_{kk'} \eqns -\frac{3i}{64\pi^3 M_{\rm p}^2}\mu (a\Omega)^4\no\\[6pt]
    &&\hspace{-1cm}\times\int^t dt'\left[ 
    \gamma_1^5 \left(\frac{m_2}{M}\right)^3e^{-i(k_x+ k'_x) x_1-i(k_y+k'_y)y_1}+\gamma_2^5 \left(\frac{m_1}{M}\right)^3
    e^{-i(k_x+ k'_x) x_2-i(k_y+k'_y)y_2}\right]\no\\[6pt]
    &&\hspace{-1.5cm}\times\frac{e^{i(\omega_k+\omega_{k'}) t'}}{\sqrt{\omega_k\omega_{k'}}}\left[\frac{\sin^2\theta}{2}+\frac{1+\cos^2\theta}{2}\cos(2\Omega t'-2\varphi)\right]\left[\frac{\sin^2\theta'}{2}+\frac{1+\cos^2\theta'}{2}\cos(2\Omega t'-2\varphi')\right]\, ,\label{eq:beta1}
     \eeqn
and
    \beqn
    \beta^{(\times)}_{kk'} \eqns -\frac{3i}{64\pi^3M_{\rm p}^2}\mu (a\Omega)^4\no\\[6pt]
    &&\times\int^t dt'\left[ 
    \gamma_1^5 \left(\frac{m_2}{M}\right)^3e^{-i(k_x+ k'_x) x_1-i(k_y+k'_y)y_1}+\gamma_2^5 \left(\frac{m_1}{M}\right)^3
    e^{-i(k_x+ k'_x) x_2-i(k_y+k'_y)y_2} \right]\no\\[6pt]
    &&\hspace{2cm}
    \times\frac{e^{i(\omega_k+\omega_{k'}) t'}}{\sqrt{\omega_k\omega_{k'}}}\cos\theta\sin(2\Omega t'-2\varphi)\cos\theta'\sin(2\Omega t'-2\varphi')\ .
    \label{eq:beta2}
    \eeqn
    
For simplicity, similar to the case of coherent states, we set
     \beqn 
    \gamma_1^5 \left(\frac{m_2}{M}\right)^3e^{-i(k_x+ k'_x) x_1-i(k_y+k'_y)y_1}+\gamma_2^5 \left(\frac{m_1}{M}\right)^3
    e^{-i(k_x+ k'_x) x_2-i(k_y+k'_y)y_2}\simeq 1\ .
    \eeqn
Based on these discussions, we can write $\beta^{(+)}$ specifically for $\bm{k}=-\bm{k}'$, that is, $(\theta',\varphi')=(\pi-\theta, \varphi+2\pi)$, as follows:
    \beqn
    \beta^{(+)}_{kk'} \eqns -\frac{3i}{64\pi^3 M_{\rm p}^2}\mu (a\Omega)^4\int^t dt' \frac{e^{i(\omega_k+\omega_{k'}) t'}}{\sqrt{\omega_k\omega_{k'}}}\no\\[6pt]
    &&\times \left[\frac{\sin^2\theta}{2}+\frac{1+\cos^2\theta}{2}\cos(2\Omega t'-2\varphi)\right]\left[\frac{\sin^2(\pi-\theta)}{2}+\frac{1+\cos^2(\pi-\theta)}{2}\cos(2\Omega t'-2(\varphi+\pi))\right]\no\\[6pt]
    \eqns -\frac{3i}{64\pi^3 M_{\rm p}^2}\mu (a\Omega)^4\int^t dt' \frac{e^{i(\omega_k+\omega_{k'}) t'}}{\sqrt{\omega_k\omega_{k'}}}\left[\frac{\sin^2\theta}{2}+\frac{1+\cos^2\theta}{2}\cos(2\Omega t'-2\varphi)\right]^2\, .
     \eeqn
Thus, the magnitude of $\beta$ is estimated 
    \beqn
    |\beta| \simeq \frac{3 \mu (a\Omega)^4}{64\pi^3 M_{\rm p}^2} \frac{1}{k^2}\ .
     \eeqn
Note that the strength of the squeeze effect depends on the direction of gravitational wave emission. To determine the squeezing parameter $\zeta$, it is necessary to convert it into a dimensionless and volume independent quantity, so it is weighted by the corresponding wave number  space volume $(k=2\Omega)$.
    \beqn
    \zeta \simeq \frac{4\pi}{3}k^3|\beta| \simeq \frac{1}{8\pi M_{\rm p}^2}\mu(a\Omega)^4 f \ ,
    \eeqn
where we used the relation $k=2\pi f$.
\printbibliography
\end{document}